\documentclass[doublecol]{epl2}
\usepackage{amssymb,amsmath,graphicx,setspace,color,rotating,subfigure,url}
\usepackage{extarrows}
\usepackage{ulem}
\usepackage{CJK}
\title{Analytic degree distributions of horizontal visibility graphs mapped from unrelated random series and multifractal binomial measures}
\shorttitle{Analytic degree distributions of horizontal visibility graphs} 

\author{Wen-Jie Xie\inst{1,2} \and Rui-Qi Han\inst{2,3} \and Zhi-Qiang Jiang\inst{1,2}\footnote{e-mail: zqjiang@ecust.edu.cn} \and Lijian Wei\inst{4} \and Wei-Xing Zhou\inst{1,2,3}\footnote{e-mail: wxzhou@ecust.edu.cn}}
\shortauthor{W.-J. Xie \etal}

\institute{

  \inst{1} Department of Finance, East China University of Science and Technology, Shanghai 200237, China\\
  \inst{2} Research Center for Econophysics, East China University of Science and Technology, Shanghai 200237, China\\
  \inst{3} Department of Mathematics, East China University of Science and Technology, Shanghai 200237, China\\
  \inst{4} School of Business, Sun Yat-Sen University, Guangzhou 510275, Guangdong, China\\
}


\abstract{
Complex network is not only a powerful tool for the analysis of complex system, but also a promising way to analyze time series. The algorithm of horizontal visibility graph (HVG) maps time series into graphs, whose degree distributions are numerically and analytically investigated for certain time series. We derive the degree distributions of HVGs through an iterative construction process of HVGs. The degree distributions of the HVG and the directed HVG for random series are derived to be exponential, which confirms the analytical results from other methods. We also obtained the analytical expressions of degree distributions of HVGs and in-degree and out-degree distributions of directed HVGs transformed from multifractal binomial measures, which agree excellently with numerical simulations.
}

\begin{document}
\maketitle



\section{Introduction}
\label{S1:Introduction}

In recent years, complex network theory has been applied to diverse fields \cite{Albert-Barabasi-2002-RMP,Newman-2003-SIAMR,Boccaletti-Latora-Moreno-Chavez-Hwang-2006-PR,Bhalla-Iyengar-1999-Science,Newman-2001-PNAS,Newman-2001a-PRE,Newman-2001b-PRE}. As a novel and powerful tool for the analysis of time series in complex systems, the theory has attracted extensive attention. Quite a few methods have been proposed to map the time series to networks, such as the temporal graph \cite{Shirazi-Jafari-Davoudi-Peinke-Tabar-Sahimi-2009-JSM,Kostakos-2009-PA}, the cycle network \cite{Zhang-Small-2006-PRL,Zhang-Sun-Luo-Zhang-Nakamura-Small-2008-PD}, the nearest neighbor network \cite{Xu-Zhang-Small-2008-PNAS}, the $n$-tuple network \cite{Li-Wang-2006-CSB,Li-Wang-2007-PA}, the recurrence network \cite{Zou-Pazo-Romano-Thiel-Kurths-2007-PRE,Marwan-Romano-Thiel-Kurths-2007-PR,Marwan-2008-EPJST,Donner-Zou-Donges-Marwan-Kurths-2010-NJP,Marwan-Donges-Zou-Donner-Kurths-2009-PLA}, the segment correlation network \cite{Yang-Yang-2008-PA}, and the visibility graph (VG) \cite{Lacasa-Luque-Ballesteros-Luque-Nuno-2008-PNAS,Lacasa-Luque-Luque-Nuno-2009-EPL,Ni-Jiang-Zhou-2009-PLA,Qian-Jiang-Zhou-2010-JPA,Yang-Wang-Yang-Mang-2009-PA,Elsner-Jagger-Fogarty-2009-GRL}.

Among these networks constructed from different methods, visibility graphs and horizontal visibility graphs (HVGs) \cite{Luque-Lacasa-Ballesteros-Luque-2009-PRE} have been widely applied in real systems, including financial markets \cite{Ni-Jiang-Zhou-2009-PLA,Qian-Jiang-Zhou-2010-JPA,Yang-Wang-Yang-Mang-2009-PA,Goncalves-Atman-2017-JNTF}, biological systems \cite{Lacasa-Luque-Luque-Nuno-2009-EPL,Shao-2010-APL,Dong-Li-2010-APL,Ahmadlou-Adeli-Adeli-2010-JNT}, ecological systems \cite{Elsner-Jagger-Fogarty-2009-GRL,Tang-Liu-Liu-2010-MPLB}, and some other complex systems \cite{Liu-Zhou-Yuan-2010-PA,Fan-Guo-Zha-2012-PA}. There are also extensions of VGs and HVGs, including limited penetrable visibility graph \cite{Zhou-Jin-Gao-Luo-2012-APS,Gao-Cai-Yang-Dang-Zhang-2016-SR,Gao-Cai-Yang-Dang-2017-PA}, binary visibility graph (BVG) \cite{Ahadpour-Sadra-2012-IS}, Markov-binary visibility graph (MBVG) \cite{Ahadpour-Sadra-ArastehFard-2014-IS}, parametric natural visibility graph (PNVG) \cite{Bezsudnov-Snarskii-2014-PA,Snarskii-Bezsudnov-2016-PRE}, multiplex horizontal visibility graph \cite{Zou-Donner-Marwan-Small-Kurths-2014-NPG,Lacasa-Nicosia-Latora-2015-SR,Bianchi-Livi-Alippi-Jenssen-2017-SR}, and directed graph due to the temporal direction of time series \cite{Lacasa-Luque-Ballesteros-Luque-Nuno-2008-PNAS,Lacasa-Nunez-Roldan-Parrondo-Luque-2012-EPJB}.

Quite a few topological properties of VGs and HVGs mapped from time series of different complex systems have been studied numerically and analytically. The main contributions are attributed to L. Lacasa and his collaborators. For instance, Iacovacci and Lacasa defined sequential motifs and derived their profiles for VGs \cite{Iacovacci-Lacasa-2016b-PRE} and HVGs \cite{Iacovacci-Lacasa-2016a-PRE}. Another topic is about the degree distribution. For random series extracted from a uniform distribution in $[0, 1]$, the degree distribution has an exponential tail $P(k)\sim e^{k/k_{0}}$ \cite{Lacasa-Luque-Ballesteros-Luque-Nuno-2008-PNAS}. For fractal time series such as fractional Brownian motions (FBMs) and Conway series, the degree distributions have power-law tails $P(k)\sim k^{-\alpha}$ \cite{Lacasa-Luque-Ballesteros-Luque-Nuno-2008-PNAS}. The tail exponent $\alpha$ for FBMs, multifractal random walks and generic $f^{-\beta}$ noises decreases linearly with the Hurst index $H$ \cite{Lacasa-Luque-Luque-Nuno-2009-EPL,Ni-Jiang-Zhou-2009-PLA}.

Our understanding about the degree distributions of HVGs is more advanced. It has been proven that the degree distribution of any HVG mapped from random series without temporal correlations has an exponential form $P(k)=(3/4)e^{-k\ln(3/2)}$ \cite{Luque-Lacasa-Ballesteros-Luque-2009-PRE,Lacasa-2014-NL}.
%
Numerical simulations show that the HVGs mapped from many chaotic time series, correlated time series and AR processes also have exponential degree distributions $P(k)\sim e^{-\lambda{k}}$, which cannot be classified by a critical value $\lambda=\ln(3/2)$ \cite{Lacasa-Toral-2010-PRE,Ravetti-Carpi-Goncalves-Frery-Rosso-2014-PLoS1,Zhang-Zou-Zhou-Gao-Guan-2017-CNSNS}.
The width of the exponential distribution decreases with the Hurst index for FBMs \cite{Xie-Zhou-2011-PA}.
N{\'{u}}{\~{n}}ez {\textit{et al.}} investigated the degree distributions of HVGs constructed from the trajectories generated by unimodal logistic maps during the type-I intermittency route to (or out of) chaos, which is close to an inverse tangent bifurcation \cite{Nunez-Luque-Lacasa-Gomez-Robledo-2013-PRE}. They found that the degree distribution is contributed by three types of nodes: $P_l(k; \epsilon) = 1/3$ when $k=2$, 3, 6 and zero otherwise for the HVG nodes belonging to the laminar phases, $P_c(k; \epsilon) \propto e^{-ck}$ for the nodes belonging to the chaotic bursts, and $P_p(k; \epsilon) \sim k^{-\alpha} \frac{\epsilon^{\alpha/2}}{|k-k_p|+1}$ where $k_p=a\epsilon^{-1/2}$ for the nodes corresponding to the peaks.
Lacasa presented a diagrammatic formalism to determine analytically the degree distribution through series expansion and provided a constructive solution for general Markovian stochastic processes and deterministic maps \cite{Lacasa-2014-NL}.
He also addressed the problem for integer-valued time series \cite{Lacasa-2016-JPA}.

In this paper, we propose an iterative approach for the analytical derivation of degree distributions of HVGs and directed HVGs (DHVGs). We derive the distributions of HVGs and DHVGs mapped from random series, which recovers the results in Refs.~\cite{Luque-Lacasa-Ballesteros-Luque-2009-PRE,Lacasa-2014-NL}. We are also able to obtain the analytical expressions of the degree distributions of HVG and DHVG mapped from multifractal binomial measures.

\section{Construction of HVG and DHVG}

\begin{figure}[t!]
\centering
\includegraphics[width=7cm]{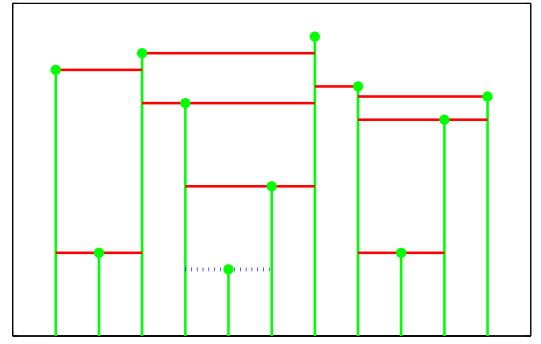}
\caption{\label{Fig:HVGAdd} (Color online.) Illustrative example of the construction of undirected HVGs mapped from time series containing 11 data points.}
\end{figure}

To have an overview of the horizontal visibility algorithm \cite{Luque-Lacasa-Ballesteros-Luque-2009-PRE}, Fig.~\ref{Fig:HVGAdd}
shows an example HVG constructed from a time series with 11 data points. Each data point is considered as a node in the HVG, two nodes will be connected if they can horizontally see each other, which means that no data points exceed the value of these two nodes.  Mathematically, we have
\begin{equation}
   x_{i}, x_{j} >  x_{n}.
    \label{Eq:HVG}
\end{equation}
for all $n$'s with $i<n<j$. We denote that $V = \{v_{i}\}$ is the set of nodes corresponding to the data in time series $\{x_{i}\}_{i = 1,...,L}$ and $E = \{e_{ij}\}$ is the adjacent matrix of graph $G= \langle V,E\rangle$. The element $e_{ij} = 1$ means that node $v_{i}$ and node $v_{j}$ are connected.
The degree of node $i$ is defined as $k_{i}=\sum_{j=1}^{L}e_{ij}$ in the undirected HVG.

\begin{figure}[t!]
\centering
\includegraphics[width=7cm]{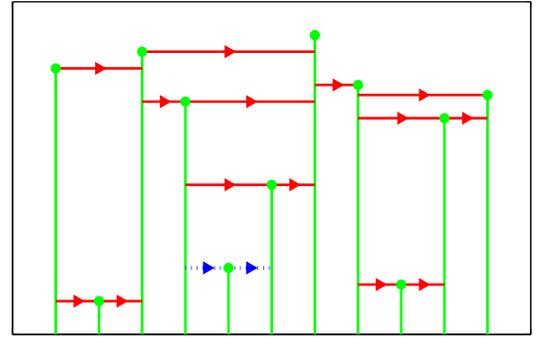}
\caption{\label{Fig:DHVGAdd} (Color online.) Illustrative example of the construction of directed HVGs mapped from time series containing 11 data points.}
\end{figure}

We can also define a directed HVG, in which the direction is time, and denote the graph as $G_{d} = \langle V_{d},E_{d}\rangle$, where $E_{d} = \{e_{ij}\}$  is the adjacent matrix of the DHVG $G_d$. $e_{ij} = 1$ means that there is a link from node $v_{i}$ to node $v_{j}$ on a given condition $i<j$.
The adjacent matrix $E_{d} = \{e_{ij}\}$ of directed graph $G_{d}$ is an upper triangular matrix which means that the adjacent matrix $E_{d}$ belongs to the upper triangular part of $E$.
Therefore, we can conclude that $k_{i} =k_{i}^{\rm{in}}+k_{i}^{\rm{out}}$ with
in-degree of node defined as $k_{i}^{\rm{in}}=\sum_{j=1}^{L}e_{ji}$ and the out-degree defined as $k_{i}^{\rm{out}}=\sum_{j=1}^{L}e_{ij}$.
Fig.~\ref{Fig:DHVGAdd} presents an example of directed HVG mapped form time series with 11 data points. In the next section, we will analyze the degree distributions of the HVGs mapped from different time series. For random time series and binomial multifractal measures, the degree distributions will be derived theoretically.

\section{Degree distribution for random series}

\subsection{Degree distribution of HVG mapped from random series}

Luque et al provided a detailed analysis and description about the degree distribution of the HVG mapped from random time series \cite{Luque-Lacasa-Ballesteros-Luque-2009-PRE}. Here
we derive alternatively the theoretical function of degree distribution based on the master equation. Generating a time series of size $L$ is equivalent to putting $L$ numbers into $L$ positions. In the first step, we randomly choose a position and put the largest number on it. In the second step, we choose a position from the rest $L-1$ positions and put the second largest number on it. In the $n$th step, we randomly choose a position from the rest $L-n+1$ positions and put the $n$th largest number on it.

In general, we can construct an HVG during the series generating process. For example, in the $n$th step, we have a series with a size of $n-1$ and we denote $N(k,n-1)$ as the number of nodes with degree $k$ in the corresponding HVG. When we put the last number into the series, the resulting HVG will have $n$ nodes. Because the last number is the smallest one, only two new edges will be added to the HVG, as shown by the dashed line segments in Fig.~\ref{Fig:HVGAdd}. And these two new edges will link to the two nodes adjacent to the $n$th node. At the same time, the degree of each of the two nodes will increase by 1. Since the new node is randomly placed, $n - 1$ nodes have the same probability $2/(n-1)$ to change their degrees. Certainly, we have to note that the $n$th node may be placed at the beginning of or at the end of the $n-1$ nodes. However, when the number $n$ is very large, the probability of such events is very small. We ignore the influence of these two ``extreme'' situations.

When a new node is added, two nodes increase by 1 in their degrees, from $k$ to $k+1$, and the degrees of the rest nodes remain unchanged. For each node, the probability of degree being changed is $2/(n-1)$ and the probability of degree remaining unchanged is $1-2/(n-1)$. Hence the number of nodes with degree $k$ in the HVG containing $n$ nodes can be calculated by the following formula:
\begin{equation}
\begin{aligned}
    N(k,n)&=\left(1-\frac{2}{n-1}\right)N(k,n-1)\\
    &+\frac{2}{n-1}N(k-1,n-1)+\delta_{k2},
    \label{Eq:Main1}
\end{aligned}
\end{equation}
where $\delta_{k2}=1$ when $k=2$ and 0 otherwise, because the degree of each new node is $k=2$.
We define the probability of nodes with degree $k$ in the HVG containing $n$ nodes as follows:
\begin{equation}
  p(k,n)=N(k,n)/n.
\end{equation}
Hence, Eq.~(\ref{Eq:Main1}) can be rewritten as follows:
\begin{equation}
    p(k,n)\approx\left(1-\frac{3}{n}\right)p(k,n-1)+\frac{2}{n}p(k-1,n-1),
    \label{Eq:p_r3}
\end{equation}
in which $\delta_{k2}/n=0$ for large $n$. When $n\rightarrow \infty$, we have

\begin{equation}
     \left\{
    \begin{aligned}
      &p(k-1,n)=p(k-1,n-1)=p(k-1),\\
      &p(k,n)=p(k,n-1)=p(k).
    \end{aligned}
    \right.
    \label{Eq:p_r4}
\end{equation}

Combining Eq.~(\ref{Eq:p_r4}) and Eq.~(\ref{Eq:p_r3}), we obtain
\begin{equation}
p(k)=\frac{2}{3}p(k-1).
    \label{Eq:p_r5}
\end{equation}
By applying $\sum_{k=2}^{\infty}p(k)=1$, we can obtain the solution of Eq.~(\ref{Eq:p_r5}):
\begin{equation}
p(k)=\frac{3}{4}\left(\frac{2}{3}\right)^k.
    \label{Eq:p_r6}
\end{equation}
This result is consistent with the analytical expression in Refs.~\cite{Luque-Lacasa-Ballesteros-Luque-2009-PRE,Lacasa-2014-NL}.

\subsection{Degree distribution of the DHVG constructed from random time series}


The degree distribution of a DHVG can also be derived. Similar to the derivation process of HVGs, we construct the directed HVG (DHVG) from the time series with $n-1$ numbers in the $n$th step. We define $N_{d}(k,n-1)$ as the number of nodes with degree (in- or out-degree) $k$. When we add the $n$th largest number into the DHVG, only two new edges will be generated shown as dashed arrows in Fig.~\ref{Fig:DHVGAdd} and the two new edges link to the two nodes adjacent to the $n$th node. The out-degree of the left node and the in-degree of the right node will increase by 1. Since the new node is placed randomly, only one endpoint node's in-degree or out-degree will increase by 1 with a probability of $1/(n-1)$. The other nodes' degrees remain the same with a probability of $1-1/(n-1)$. So the number of nodes with degree $k$ in the new DHVG containing $n$ nodes can be calculated by the formula as follows:
\begin{equation}
\begin{aligned}
    N_{d}(k,n)&=\left(1-\frac{1}{n-1}\right)N_{d}(k,n-1)\\
    &+\frac{1}{n-1}N_{d}(k-1,n-1)+\delta_{k1},
    \label{Eq:directedMain1}
\end{aligned}
\end{equation}
where $\delta_{k1}=1$ when $k=1$ and 0 otherwise, because the degree of each new node is $k=1$.
If $k=1$, we have $\delta_{k1}=1$. When $k\neq 1$, we have $\delta_{k1}=0$.
We define $p_d(k,n)=N_{d}(k,n)/n$ as the probability of nodes with the
in-degree or out-degree $k$ in the DHVG containing $n$ nodes. So Eq.~(\ref{Eq:directedMain1}) can be rewritten as follows
\begin{equation}
    p_{d}(k,n)=\left(1-\frac{2}{n}\right)p_{d}(k,n-1)+\frac{1}{n}p_{d}(k-1,n-1).
    \label{Eq:directedp_r3}
\end{equation}
When $n\rightarrow \infty$, we have

\begin{equation}
     \left\{
    \begin{aligned}
      &p_{d}(k-1,n)=p_{d}(k-1,n-1)=p_{d}(k-1),\\
      &p_{d}(k,n)=p_{d}(k,n-1)=p_{d}(k).
    \end{aligned}
    \right.
    \label{Eq:directedp_r4}
\end{equation}

Combining Eq.~(\ref{Eq:directedp_r4}) and Eq.~(\ref{Eq:directedp_r3}), we obtain
\begin{equation}
p_{d}(k)=\frac{1}{2}p_{d}(k-1).
    \label{Eq:directedp_r5}
\end{equation}
By applying $\sum_{k=1}^{k=\infty}p_{d}(k)=1$, we can obtain the solution of Eq.~(\ref{Eq:directedp_r5}):
\begin{equation}
p_{d}(k)=\left(\frac{1}{2}\right)^k.
    \label{Eq:directedp_r6}
\end{equation}
This result is also consistent with the analytical expression in Refs.~\cite{Luque-Lacasa-Ballesteros-Luque-2009-PRE,Lacasa-2014-NL}.

\section{Degree distribution for multifractal binomial measures}

Based on the $p$-model \cite{Meneveau-Sreenivasan-1987-PRL}, we generate multifractal binomial measures with length $2^m$. When $m=1$,  the length of time series is 2 and the two data points are $p_1$ and $p_2=1-p_1$. When $m=2$, the length of time series is 4 and the four data points are $p_1^2$, $p_1p_2$, $p_1p_2$, $p_2^2$.
We can represent the multifractal binomial measures with length $2^m$ as series $\{p_1^{k_i}(1-p_1)^{m-k_i}\}_{i = 1,...,2^m}$, where $k_{i}\in\{0,1,2,...,m\}$. The time series can be rewritten as $\{[p_1/(1-p_1)]^{k_i}(1-p_1)^{m}\}_{i = 1,...,2^m}$. When $p_1>0.5$, we have $p_1/(1-p_1)>1$. $[p_1/(1-p_1)]^{k_i}$ is a monotonically increasing function of the  independent variable $k_i$. The size relationship among the numbers depend son the index $k_i$. So the time series can be rewritten as $\{k_i\}_{i = 1,...,2^m}$.
The DHVGs mapped from binomial measures do not depend on the multiplier $p_1$ for $1>p_1>0.5$. We can conduct the same analysis for $p_1<0.5$ and the DHVGs mapped from binomial measures do not depend on the multiplier $p_1$ for $0<p_1<0.5$.
It should be noted that the multifractal binomial measures with $p_1=p$ is the reverse order of the multifractal binomial measures with $p_1=1-p$, that is, the HVGs mapped from binomial measures do not depend on the multiplier $p_1$ for $0<p_1<1$ and $p_1\neq0.5$.
What is more, we can get the DHVG mapped from the multifractal binomial measures with $p_1=p$ by flipping the directions of all links in DHVG mapped from the multifractal binomial measures with $p_1=1-p$.
Hence the out-degree (in-degree) distribution of the DHVG constructed from multifractal binomial measures with $p_1=p$ is the same with the in-degree (out-degree) distribution of the DHVG constructed from multifractal binomial measures with $p_1=1-p$.
In our simulations, we use parameters $p_1=0.25$ and $p_2=1-p_1=0.75$.

\subsection{Degree distribution of the HVG constructed from multifractal binomial measures}

\begin{figure}[htb]
  \centering
  \includegraphics[width=7cm]{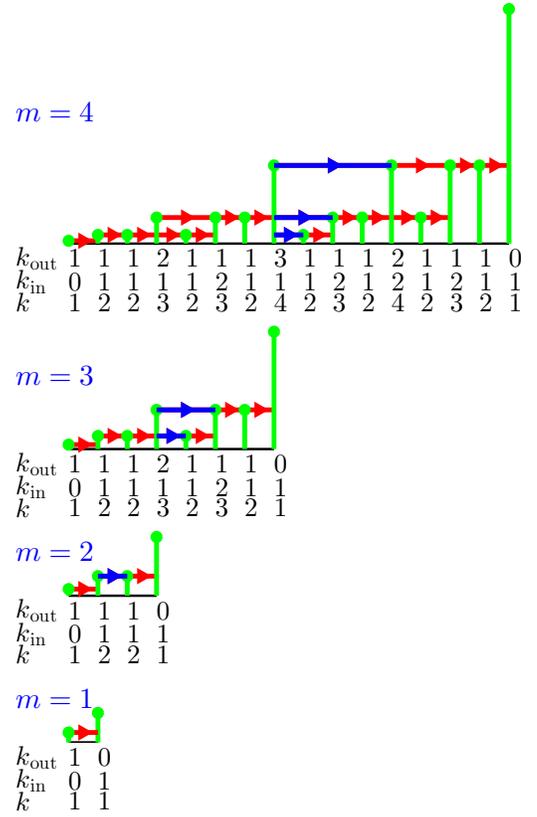}
  \caption{\label{Fig:HVG_Plot_pmodel_Dout} Illustrative example showing the construction of HVGs and DHVGs mapped from multifractal binomial measures.}
\end{figure}

As shown in Fig.~\ref{Fig:HVG_Plot_pmodel_Dout},  the HVG containing $2^m$ nodes consists of two HVGs with $2^{m-1}$ nodes as well as $m-1$ edges. The number of edges in the HVG with $2^m$ nodes can be calculated as
\begin{equation}
E(m)=2E(m-1)+m-1.
 \label{Eq:pmodel_E1}
\end{equation}
We can transform the above equation into geometric series such as
$E(m)+m+1=2[E(m-1)+(m-1)+1]$.
Consider the geometric series $b_m=2\times b_{m-1}$, where $b_m=E(m)+m+1$ and $b_1=3$. It is easy to obtain the number of edges in the HVG mapped from multifractal binomial measures of length $2^m$:
\begin{equation}
E(m)=3\times2^{m-1}-m-1.
 \label{Eq:pmodel_E2}
\end{equation}

We denote $N(1,2^m)$ as the number of nodes with degree $k=1$ in the HVG containing~$2^m$~nodes. In Fig.~\ref{Fig:HVG_Plot_pmodel_Dout}, $N(1,2^m)=2$ nodes locate at the start and the end of the time series. The HVG with $2^m$ nodes consists of two HVGs containing $2^{m-1}$ nodes and $m-1$ edges. The $m-1$ edges added into the graph are illustrated in blue color in Fig.~\ref{Fig:HVG_Plot_pmodel_Dout}. Moreover, the $m-1$ new edges are connected to the $2^{m-1}$th node, which are shown at the end of the left half of the time series. The $2^{m-1}$th node's degree increases from $1$ to $m$. At the same time, the degrees of the $m-1$  nodes at the right half of the time series increase by 1, and what is more, the degrees of the $m-1$  nodes are 1, 2, 3, $\cdots$, $m-1$, respectively. Let $k=2, 3, \cdots, m-1$, we can obtain the following equation:
\begin{equation}
N(k,2^m)=2N(k,2^{m-1}),~~~~1<k<m
 \label{Eq:pmodel_Skm}
\end{equation}
If $k=m$, the nodes with degree $k=m$ are composed of two nodes. The first node is the $2^{m-1}$th node, whose degree increases from 1 to $m$, while the second node is the one whose degree increases from $m-1$ to $m$. Hence, the number of nodes with degree $k=m$ is $N(m,2^m)=2$. We obtain the number of nodes with degree $k$ in the $m$th iteration as follows:
\begin{equation}
     N(k,2^m)=\left\{
    \begin{aligned}
      &2,~~~~~~~~~~~~k=1\\
      &2^{m-k+1},~~~1<k\leq m\\
    \end{aligned}
    \right.
    \label{Eq:pmodel_Smm_all}
\end{equation}
To calculate the degree distribution of the HVG with $2^m$ nodes, we use the formula $p(k,2^m)=N(k,2^m)/2^m$ and rewrite Eq.~(\ref{Eq:pmodel_Smm_all}) as follows
\begin{equation}
     p(k,m)=\left\{
    \begin{aligned}
      &\left({1}/{2}\right)^{m-1},~~~~~~k=1\\
      &\left({1}/{2}\right)^{k-1},~~~1<k\leq m
    \end{aligned}
    \right.
    \label{Eq:Pmodel:HVG:PDF:k}
\end{equation}


\subsection{Out-degree distribution of the DHVG constructed from multifractal binomial measures}

We denote $N_{\rm{out}}(k,2^m)$ as the number of nodes with the out-degree $k_{\rm{out}}=k$ in the DHVG with $2^m$ nodes. Based on the DHVG algorithm, all the directed edges are from the left to right. According to Fig.~\ref{Fig:HVG_Plot_pmodel_Dout}, there is only one node with $k_{\rm{out}}=0$, which is at the right end of the time series. Hence, we have $N_{\rm{out}}(0,2^m)=1$.

At the $m$-step, the length of the time series grows from $2^{m-1}$ to $2^m$ and $m-1$ new edges are connected to the $2^{m-1}$th node (see blue arrows in Fig.~\ref{Fig:HVG_Plot_pmodel_Dout}). The $2^{m-1}$th node's out-degree increases from $k_{\rm{out}}=0$ to $k_{\rm{out}}=m-1$.
The out-degrees of all other nodes from 1 to $2^{m-1}-1$ remain unchanged. Due to the self-similarity of the binomial measure, the out-degrees of the ``new'' nodes from $2^{m-1}+1$ to $2^m$ are the same as those in the $(m-1)$-step DHVG. Therefore, we have
\begin{equation}
 N_{\rm{out}}(k,2^m)=2N_{\rm{out}}(k,2^{m-1}),~~{\mathrm{for}}~~0<k<m-1,
 \label{Eq:pmodel_Soutkm}
\end{equation}
where $m>2$. When $m=2$, the number of nodes with out-degree $k_{\rm{out}}=1$ is $N_{\rm{out}}(1,2^2)=3$. Inserting $N_{\rm{out}}(1,2^2)=3$ into Eq.~(\ref{Eq:pmodel_Soutkm}), we get
\begin{equation}
 N_{\rm{out}}(1,2^m)=3\times2^{m-2},~~{\mathrm{for}}~~m\geq2.
 \label{Eq:pmodel_Sout1m2k1}
\end{equation}
When $m=3$, the number of nodes with out-degree $k_{\rm{out}}=2$ is $N_{\rm{out}}(2,2^3)=1$. Inserting $N_{\rm{out}}(2,2^3)=1$ into Eq.~(\ref{Eq:pmodel_Soutkm}), we obtain
\begin{equation}
 N_{\rm{out}}(2,2^m)=2^{m-3},~~{\mathrm{for}}~~m\geq3.
 \label{Eq:pmodel_Sout1m3k2}
\end{equation}
Similarly, we can obtain
\begin{equation}
  N_{\rm{out}}(k,2^m)=2^{m-k-1},~~{\mathrm{for}}~~m\geq{k+1}.
 \label{Eq:pmodel_Sout1m}
\end{equation}
According to Fig.~\ref{Fig:HVG_Plot_pmodel_Dout}, we can observe that the maximum out-degree is $k_{\rm{out}}=m-1$, corresponding to the $2^{m-1}$th node. In other words, $N_{\rm{out}}(m-1,2^m)=1$, which has been included in Eq.~(\ref{Eq:pmodel_Sout1m}).

Summarizing the results above, we have the formula for the number of nodes with out-degree $k$ as follows:
\begin{equation}
     N_{\rm{out}}(k,2^m)=\left\{
    \begin{aligned}
      &1,   ~~~~~~~~~~~k=0\\
      &3\times2^{m-2},~k=1\\
      &2^{m-k-1},   ~~~1<k\leq m-1
    \end{aligned}
    \right.
    \label{Eq:pmodel_Soutm_all}
\end{equation}

In order to calculate the out-degree distribution of the DHVG with $2^m$ nodes, we use the definition $p_{\rm{out}}(k,2^m)= N_{\rm{out}}(k,2^m)/2^m$ and rewrite the
Eq.~(\ref{Eq:pmodel_Soutm_all}) as follows
\begin{equation}
     p_{\rm{out}}(k,2^m)=\left\{
    \begin{aligned}
      &\left(\frac{1}{2}\right)^{m},~~~~k=0\\
      &~~~\frac{3}{4},~~~~~~~~~k=1\\
      &\left(\frac{1}{2}\right)^{k+1},~~1<k\leq m-1
    \end{aligned}
    \right.
    \label{Eq:pmodel:DHVG:PDF:kout}
\end{equation}

\subsection{In-degree distribution of the DHVG constructed from the multifractal binomial measures}

We denote $N_{\rm{in}}(k,2^m)$ as the number of nodes with in-degree $k_{\rm{in}}=k$ in the DHVG with $2^m$ nodes. In Fig.~\ref{Fig:HVG_Plot_pmodel_Dout}, all the edges are in the left-to-right direction. There is only one node with $k_{\rm{in}}=0$, which is at the start of the time series. Hence, we have $N_{\rm{in}}(0,2^m)=1$.

When the time series expands from $2^{m-1}$ data points to $2^m$ data points, $m-1$ new edges are connected from the $2^{m-1}$th node to $m-1$ nodes $2^{m-1}+2^{m'-1}$ with $m'=1, 2, \cdots, m-1$ in the right half of time series (see the blue arrows in Fig.~\ref{Fig:HVG_Plot_pmodel_Dout}). Each of these $m-1$ node's in-degree will increase by 1, when compared with their counterparts in the left half, that is,
\begin{equation}
  k_{\rm{in}}(2^{m-1}+2^{m'-1})=k_{\rm{in}}(2^{m'-1})+1,
\end{equation}
where $m'=1, 2, \cdots, m-1$. We obtain that the in-degree of node $2^{m-1}+1$ is 1 ($k_{\rm{in}}=0$ for node $2^0$) and the degrees of the other $m-2$ nodes are $2$ ($k_{\rm{in}}=1$ for nodes at $2^{m'-1}$). Therefore, we have
\begin{equation}
N_{\rm{in}}(1,2^m)=2N_{\rm{in}}(1,2^{m-1})-(m-2)+1
 \label{Eq:pmodel_Sin1m}
\end{equation}
and
\begin{equation}
N_{\rm{in}}(2,2^m)=2N_{\rm{in}}(2,2^{m-1})+(m-2).
\label{Eq:pmodel_Sin2m}
\end{equation}
It follows from Eq.~(\ref{Eq:pmodel_Sin1m}) that
\begin{equation}
N_{\rm{in}}(1,2^m)-m+1=2[N_{\rm{in}}(1,2^{m-1})-(m-1)+1].
 \label{Eq:pmodel_Sin1m1}
\end{equation}
Defining $b_m=N_{\rm{in}}(1,2^m)-m+1$, we have $b_m=2\times b_{m-1}$, where $b_2=2$. The number of nodes with $k_{\rm{in}}=1$ is obtained as follows:
\begin{equation}
N_{\rm{in}}(1,2^m)=2^{m-1}+m-1.
\label{Eq:pmodel_Sin1m2}
\end{equation}

Similarly, according to Eq.~(\ref{Eq:pmodel_Sin2m}), we have
\begin{equation}
N_{\rm{in}}(2,2^m)+m=2[N_{\rm{in}}(2,2^{m-1})+(m-1)].
 \label{Eq:pmodel_Sin2m1}
\end{equation}
Defining $b_m=N_{\rm{in}}(2,2^m)+m$, we have $b_m=2\times b_{m-1}$, where $b_3=4$. The number of nodes with $k_{\rm{in}}=2$ is obtained as follows:
\begin{equation}
N_{\rm{in}}(2,2^m)=2^{m-1}-m.
\label{Eq:pmodel_Sin2m2}
\end{equation}

\begin{figure*}[t!]
\centering
 \includegraphics[width=15cm]{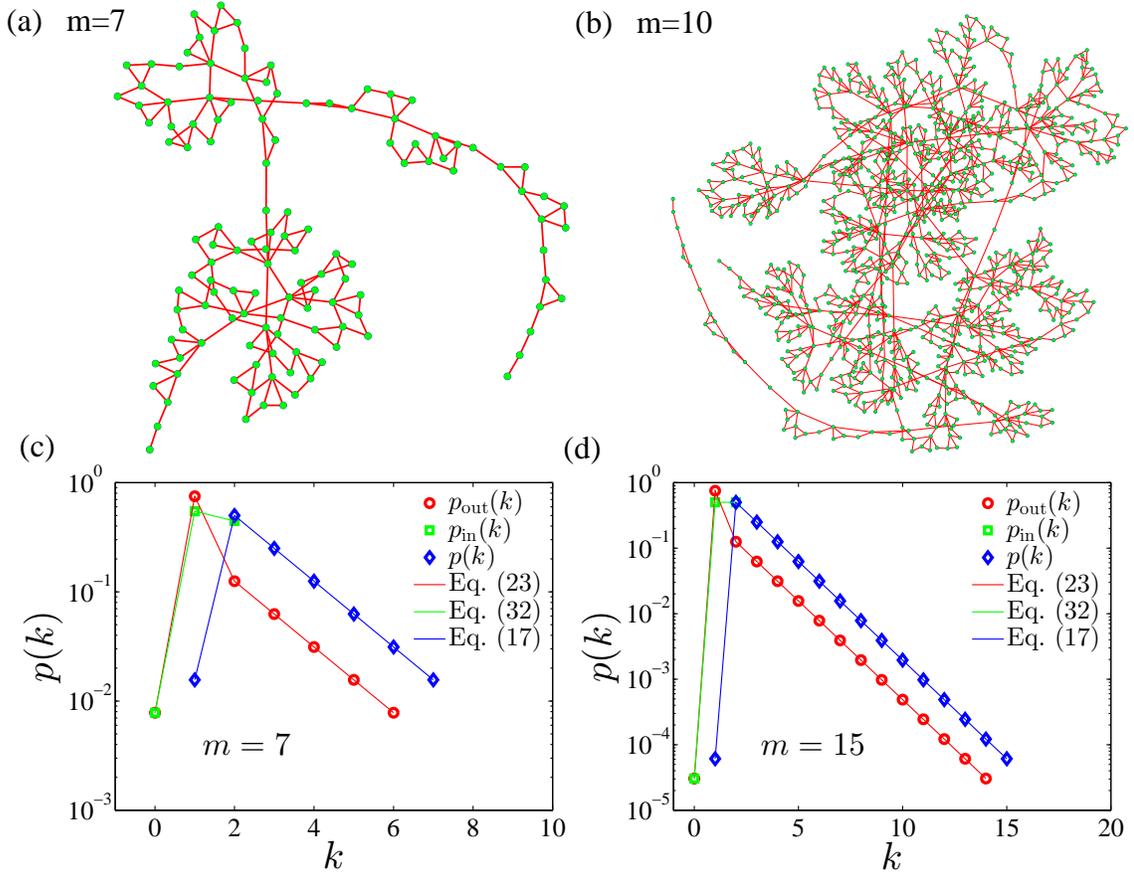}
  \caption{(a) The HVG mapped from the multifractal binomial measure with $m=7$. (b) The HVG mapped from the multifractal binomial measure with $m=10$. (c) Degree distributions of the HVG and DHVG for $m=7$. (d) Degree distributions of the HVG and DHVG for $m=15$.}
    \label{Fig:HVG:Numerical:Pk:PM}
\end{figure*}

For binomial measures, it is easy to prove that the maximum in-degree is 2. Therefore, we reach the following:
\begin{equation}
     N_{\rm{in}}(k,2^m)=\left\{
    \begin{aligned}
      &1,~~~~~~~~~~~~~~~~~~k=0\\
      &2^{m-1}+m-1,~k=1\\
      &2^{m-1}-m,~~~~~~k=2
    \end{aligned}
    \right.
    \label{Eq:pmodel_Sin2m3}
\end{equation}
The in-degree distribution of the DHVG with $2^m$ nodes can be presented as follows with the formula $p_{\rm{in}}(k,2^m)= N_{\rm{in}}(k,2^m)/2^m$
\begin{equation}
     p_{\rm{in}}(k,2^m)=\left\{
    \begin{aligned}
      &\frac{1}{2^m},~~~~~~~~~~~k=0\\
      &\frac{1}{2}+\frac{m-1}{2^m},~k=1\\
      &\frac{1}{2}-\frac{m}{2^m},~~~~~k=2
    \end{aligned}
    \right.
    \label{Eq:pmodel:DHVG:PDF:kin}
\end{equation}


\subsection{Numerical validation}


We now verify the theoretical degree distributions expressed in Eq.~(\ref{Eq:Pmodel:HVG:PDF:k}), Eq.~(\ref{Eq:pmodel:DHVG:PDF:kout}) and Eq.~(\ref{Eq:pmodel:DHVG:PDF:kin}) for the HVGs and DHVGs mapped from multifractal binomial measures through numerical experiments, which can be done by generating different iterations $m$ of time series. In Fig.~\ref{Fig:HVG:Numerical:Pk:PM} (a) and (b), we illustrate the HVGs for $m=7$ and $m=10$, respectively. It can be seen that the graphs exhibit self-similarity.

The numerically obtained degree distributions $P(k)$, $P(k_{\rm{out}})$ and $P(k_{\rm{in}})$ for $m=7$ and $m=15$ are illustrated respectively in Fig.~\ref{Fig:HVG:Numerical:Pk:PM} (c) and (d). Concerning the in-degree distribution of the DHVG for each $m$, there are only three separate data points at $k=0$, 1 and 2. The out-degree distributions of the DHVGs overlap for different $m$ values for overlapping positive $k_{\rm{out}}$ values and are exponential when $k_{\rm{out}}>1$. The degree distributions of the undirected HVGs also overlap for different $m$ values for overlapping positive $k$ values. Overall, the numerical results are in excellent agreement with the analytical results presented in Eq.~(\ref{Eq:Pmodel:HVG:PDF:k}), Eq.~(\ref{Eq:pmodel:DHVG:PDF:kout}), and Eq.~(\ref{Eq:pmodel:DHVG:PDF:kin}).

\section{Conclusion}

In this paper, we proposed a novel method for deriving degree distributions of HVGs through an iterative construction process of HVGs and illuminated the method by applying it to uncorrelated random time series and multifractal binomial measures. For random series, the iterative construction process proceeds the data in an ascending way. For binomial measure, the process iterates by expanding the measure from $m$ steps to $m+1$ steps based on the restrict self-similarity of the measures. Obviously, the HVGs mapped from binomial measures do not depends on the multiplier $p$.

The degree distributions of the HVG and the directed HVG for random series have been obtained analytically to be exponential, which confirms the analytical results from other methods in the literature. We also obtained the analytical degree distributions of HVGs and in-degree and out-degree distributions of DHVGs transformed from multifractal binomial measures. Results from numerical simulations confirmed the analytical results.

There are also some open problems unsolved in this work. The main question is about the generality of the proposed method. For more complicated time series, such as fractional Brownian motions, logistic maps and Lorenz equations, it is not clear if the proposed approach can be applied or not. Nevertheless, we believe that the approach can be applied to other time series, especially when the time series can be constructed in a self-similar way. Further studies are necessary.

\acknowledgments

This work was supported by National Natural Science Foundation of China (71532009) and Fundamental Research Funds for the Central Universities (222201718006).


\end{document}